\documentclass[12pt]{article}
\usepackage{epsfig}

\usepackage{hyperref}
\usepackage[parfill]{parskip}  
\usepackage{graphicx}
\usepackage{epstopdf}
\usepackage[T1]{fontenc}
\usepackage[absolute,overlay]{textpos}
\usepackage{fancybox}
\usepackage[dvipsnames]{xcolor}
\usepackage{amsmath, amsthm, amssymb,amsfonts}
\usepackage{caption}
\usepackage{subcaption}
\usepackage{float}
\def\Title#1{\begin{center} {\Large {\bf #1} } \end{center}}

\begin{document}

\Title{NOvA Short-Baseline Tau Neutrino Appearance Search}

\bigskip\bigskip


\begin{raggedright}  

{\it Rijeesh Keloth\index{Keloth, R.}\\
Department of Physics\\
Cochin University of Science and Technoloy\\
Cochin, Kerala 682022, INDIA\\
\bigskip\bigskip
\it Adam Aurisano, Alexander Sousa\index{Aurisano, A.}\\
Department of Physics\\
University of Cincinnati\\
Cincinnati, OH 45220, USA\\
\bigskip\bigskip
\it Gavin S Davies\index{Davies, D.}\\
Department of Physics\\
Indiana University\\
Bloomington, IN 47405, USA\\
\bigskip\bigskip
\it Louise Suter, Robert K Plunkett\index{Suter, L.}\\
Neutrino Division\\
Fermi National Accelerator Laboratory\\
Batavia, IL 60510, USA\\
}
\end{raggedright}

\begin{center}
Talk presented at the APS Division of Particles and Fields Meeting (DPF 2017), July 31-August 4, 2017, Fermilab. C170731
\end{center}

\section{Introduction}
Standard three-flavor neutrino oscillations have been well explained by a wide range of neutrino experiments. However anomalous results, such as the electron-antineutrino excess seen by LSND and MiniBooNE do not fit the three-flavor paradigm. This can be explained by an additional fourth neutrino at a larger mass scale than the existing three flavor neutrinos. The NOvA experiment consists of two finely segmented, liquid scintillator detectors operating  14.6 mrad off-axis from the NuMI muon-neutrino beam. The Near Detector is located on the Fermilab campus, 1 km from the NuMI target, while the Far Detector is located at Ash River, MN, 810 km from the NuMI target. The NOvA experiment is primarily designed to measure electron-neutrino appearance at the Far Detector using the Near Detector to control systematic uncertainties; however, the Near Detector is well suited for searching for anomalous short-baseline oscillations. We present a novel method for selecting tau neutrino interactions with high purity at the Near Detector using a convolutional neural network. Using this method, the sensitivity to anomalous short-baseline tau-neutrino appearance due to sterile neutrino oscillations will be discussed.

\section{NOvA Experiment}
The NOvA experiment consists of finely segmented  liquid scintillator Far and Near Detectors operating 14.6 mrad off-axis from the upgraded NuMI muon-neutrino beam. The Near Detector(ND) has a 0.3 kton mass and is located at the Fermilab campus, 1 km from the NuMI target. The Far Detector(FD) has a 14 kton mass and is located at Ash River, MN, 810 km from the NuMI target. The neutrino beam is produced using 120 GeV protons incident on a 1.2 m long graphite target. The kaons and pions emerging from the target are focused by two magnetic horns and either decay in flight into neutrinos over a distance of 705 m, including a 675 m decay pipe, or are absorbed. The resulting neutrino beam has a narrow energy spectrum peaked at 2 GeV. The beam is extracted every 1.33 s using a time window of 10 $\mu$s. 
%
%

\section{Searching  for the Sterile Neutrinos using Tau Neutrino Appearance}
The electron-antineutrino and neutrino excesses seen by LSND \cite{LSND} and MiniBooNE \cite{MiniBooNE}\cite{MiniBooNE1} do not fit the three-flavor paradigm. These excesses can be explained by adding a sterile neutrino at a larger mass scale than the three active neutrinos. The extended mixing matrix when adding in this extra neutrino is
\begin{equation}
\left(
\begin{matrix}
\nu_{e} \\
\nu_{\mu} \\
\nu_{\tau} \\
\nu_{s} 
\end{matrix} \right) = 
\left(
\begin{matrix}
\text{U}_{e1} & \text{U}_{e2} & \text{U}_{e3}& \text{U}_{e4} \\
\text{U}_{\mu1} & \text{U}_{\mu2} & \text{U}_{\mu3}& \text{U}_{\mu4} \\
\text{U}_{\tau1} & \text{U}_{\tau2} & \text{U}_{\tau3}& \text{U}_{\tau4} \\
\text{U}_{s1} & \text{U}_{s2} & \text{U}_{s3}& \text{U}_{s4} 
\end{matrix} \right)
\left(
\begin{matrix}
\nu_{1} \\
\nu_{2} \\
\nu_{3} \\
\nu_{4} 
\end{matrix} \right)
\end{equation}
where $\nu_{\mu}$, $\nu_{e}$, $\nu_{\tau}$, $\nu_{s}$ and $\nu_{1}$, $\nu_{2}$, $\nu_{3}$, $\nu_{4}$ are the flavor and mass eigen states respectively, and U$_{\alpha4}$, $\alpha=e, \mu, \tau, s$ represents the mixing between active and sterile neutrino. \\
The oscillation probability can be approximated for the Short-baseline oscillations as follows,
\begin{equation}
\text{P}_{\alpha \rightarrow \beta} = \delta_{\alpha\beta} - 4\sum_{i > j}\mathcal{R}(\text{U}^{*}_{\alpha i}\text{U}_{\beta i}\text{U}_{\alpha j}\text{U}^{*}_{\beta j})\text{sin}^{2}\left(\frac{\Delta m^{2}_{ij}L}{4E}\right)  +  2\sum_{i > j}\mathcal{I}(\text{U}^{*}_{\alpha i}\text{U}_{\beta i}\text{U}_{\alpha j}\text{U}^{*}_{\beta j})\text{sin}^{2}\left(\frac{\Delta m^{2}_{ij}L}{4E}\right)
\label{eq:3prob}
\end{equation}
\noindent Similarly, $\nu_{\mu} \rightarrow \nu_{\tau}$  appearance probability from eq.[\ref{eq:3prob}] can be written as
\begin{equation}\label{eqn:disapp_prob}
\text{P}^{\text{SBL},3+1}_{\overset{(-)}{\nu_{\mu}}\rightarrow \overset{(-)}{\nu_{\tau}}} =  \sin^2{2\theta_{\mu\tau}}\sin^2{\frac{\Delta m^2_{41}L}{4E}}
\end{equation} 
where $\sin^2{2\theta_{\mu\tau}} \equiv 4|\text{U}_{\mu4}|^{2}|\text{U}_{\tau4}|^{2}$ = $\cos^{4}\theta_{14} \sin^{2}2\theta_{24}\sin^{2}\theta_{34}$

\noindent and, $\nu_{\mu} \rightarrow \nu_{\mu}$ disappearance probability from eq.[\ref{eq:3prob}] can be written as
\begin{equation}\label{eqn:disapp_prob}
\text{P}^{\text{SBL},3+1}_{\overset{(-)}{\nu_{\mu}}\rightarrow \overset{(-)}{\nu_{\mu}}} = 1 - \sin^2{2\theta_{\mu\mu}}\sin^2{\frac{\Delta m^2_{41}L}{4E}}
\end{equation} 

where $\sin^2{2\theta_{\mu\mu}} \equiv 4|\text{U}_{\mu4}|^{2}(1-|\text{U}_{\mu4}|^{2})$  = $\cos^{2}\theta_{14} \sin^{2}\theta_{24}$
\begin{figure}
\centering
\begin{subfigure}{.5\textwidth}
  \centering
\epsfig{file=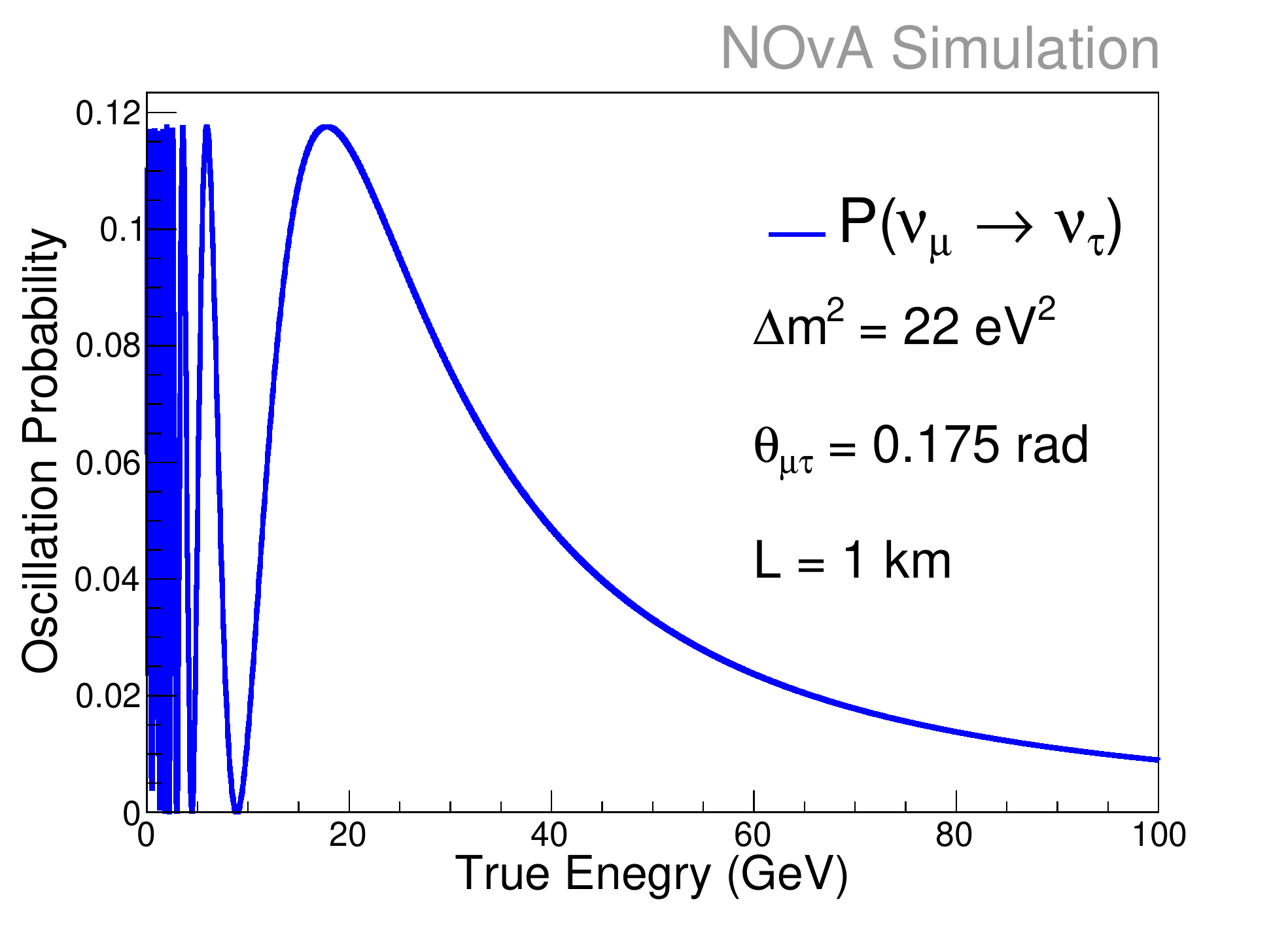,height=2.2in}
   \caption{Probability for $\nu_\mu\rightarrow\nu_\tau$ oscillations}
  \label{fig:sub1}
\end{subfigure}%
\begin{subfigure}{.5\textwidth}
  \centering
\epsfig{file=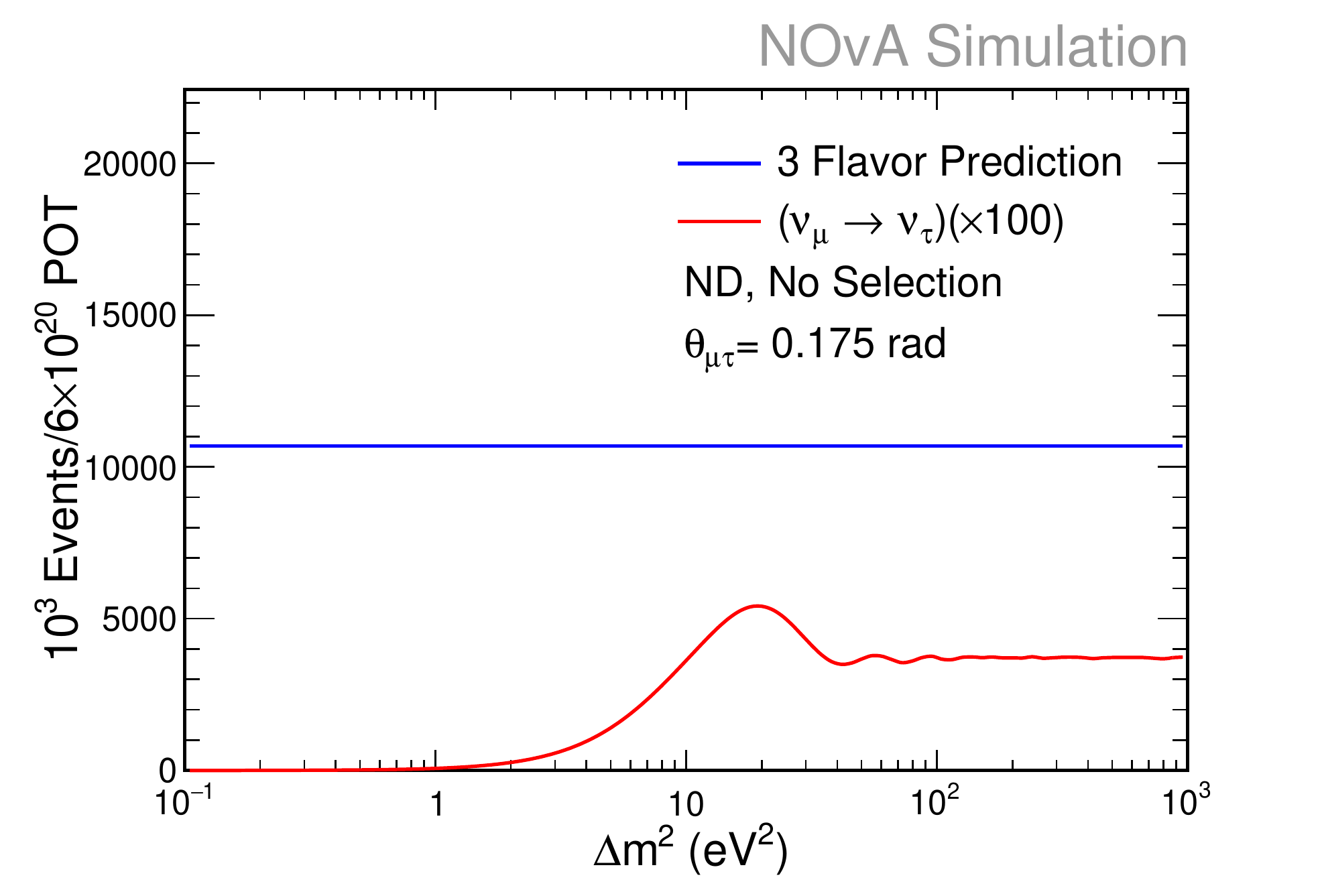,height=2.2in}
  \caption{ND event rates for  different $\Delta m^2$ values(before selection)}
  \label{fig:sub2}
\end{subfigure}
\caption{Oscillation probability(left) and event rates for different  $\Delta m^2$ values(right).}
\label{fig:test}
\end{figure}

\section{Tau Neutrino Events in NOvA ND}
Charged Current(CC) tau-neutrino interactions in the ND occur primarily between 10 and 20 GeV as you can see in Fig. 1.b, so they typically have high multiplicity hadronic systems originating from nuclear scattering in addition to the decay products of the outgoing tau-lepton in the final state. Hence we can categorize CC tau neutrino interaction as either hadronic with one outgoing neutrino and typically one or three pions, or leptonic with two outgoing neutrinos and an electron or muon.  All events are likely to have a large amount of additional hadronic activity due to the hadronic recoil associated with these high energy events. An example hadronic CC tau-neutrino interaction is shown in Fig. 2
 
 We developed the Convolutional Visual Network (CVN) \cite{Aurisano}using the tools from the computer vision community to classify neutrino interactions, which is based on the GoogLeNet convolutional neutral network (CNN) architecture. CNNs treat each event as an image, and pass these images through layers consisting of banks of learned filters to extract the features of these images. These features are then used to classify events according to neutrino flavor and interaction type. Currently CVN is also used in the electron neutrino appearance \cite{NOvA_nue_2017} and neutral-current (NC) disappearance analyses in NOvA \cite{NOvA_nc_2016}. 
 \begin{figure}[H]
\begin{center}
\epsfig{file=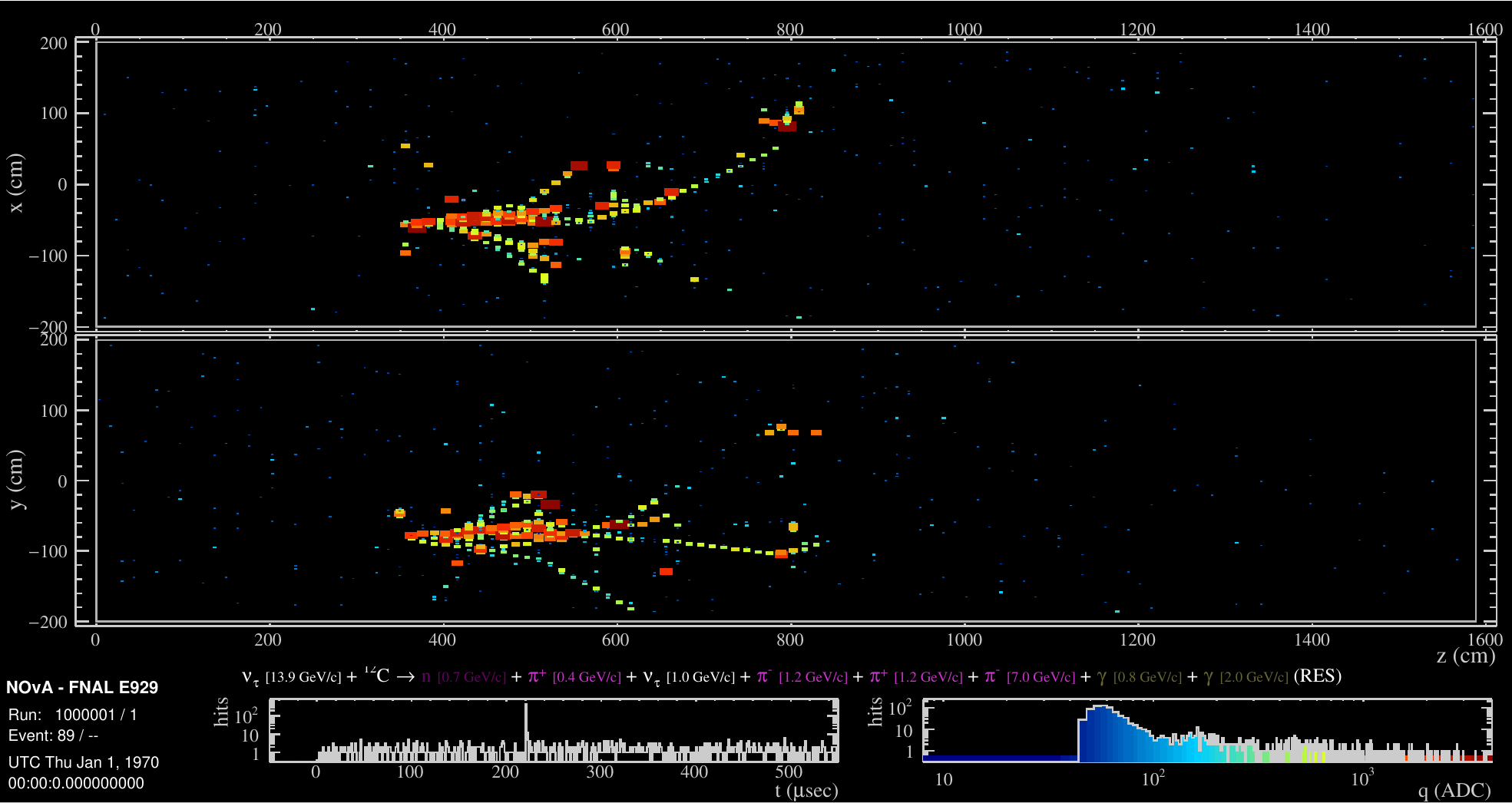,height=2.5 in}
\caption{A simulated 13.9 GeV hadronic CC tau neutrino event in NOvA ND. Color indicates the charge deposited in the calorimeter. Upper and lower plot are the top and side views, respectively of this event in the detector.}
\label{fig:magnet}
\end{center}
\end{figure}

\section{Signal Selection and Prediction}

Backgrounds arise from both misidentified NC, $\nu_{\mu}$ CC and $\nu_e$ CC interactions and from external sources. In the background NC neutrino interactions, the neutrino leaves the detector with reduced energy and products of nuclear fragmentation remain behind. This hadronic recoil appears in the detector as an isolated cluster of energy deposits which may be mimicking hadronic $\nu_{\tau}$ CC interaction. NuMI beam $\nu_{\mu}$ CC and $\nu_e$ CC events, typically with high momentum transfer to the hadronic system, can be produced where the lepton may be mimicking a leptonic CC tau neutrino interaction. The major external  source of the background are the NuMI beam events interacting in the periphery of the ND and in the surrounding cavern. 

We use the tau-neutrino CVN as the primary classifier for the signal selection and use a pre-selection which further improve the rejection of detector external activities caused by events in rock around the detector. The pre-selection include a fiducial volume cut, which selects only the events having the reconstructed neutrino interaction vertex in a restricted region inside the detector and a containment cut, which rejects the events in rock around the detector.

This analysis applies a joint fit for the parameters  $\Delta m^2_{14}$ and $\theta_{\mu\tau}$ using $\nu_\mu\rightarrow\nu_\tau$  and $\nu_\mu\rightarrow\nu_\mu$  selections in ND. A high energy $\nu_\mu\rightarrow\nu_\mu$ selection helps to constrain the highly correlated systematic uncertainties in the analysis.
Table.1 shows the number of predicted neutrino events for $\nu_\mu\rightarrow\nu_\mu$  and $\nu_\mu\rightarrow\nu_\tau$  selections at the point $\Delta m^2_{14}$ = 22 e$V^2$ and $\theta_{\mu\tau}$ = 0.175 rad. for 18$\times10^{20}$ POT.
\begin{table}[H]
\begin{center}
\begin{tabular}{l|cccc}  
Selection                                                    &  $\nu_\mu$ CC                      & NC            &  $\nu_e$ CC          & $\nu_\tau$ CC\\ \hline
 $\nu_\mu\rightarrow\nu_\mu$                   &   3.31$\times10^6$                                     &     32072      &     4637                 & 14955 \\
$\nu_\mu\rightarrow\nu_\tau$    &  336                                     &     297     & 177 			    & 612 \\ \hline
\end{tabular}
\caption{ The neutrino events passing $\nu_\mu\rightarrow\nu_\mu$  and $\nu_\mu\rightarrow\nu_\tau$  selections at the point $\Delta m^2_{14}$ = 22 e$V^2$ and $\theta_{\mu\tau}$ = 0.175 rad.  for 18$\times10^{20}$ POT.}
\label{tab:blood}
\end{center}
\end{table}

 Fig. 3(a) and 3(b) show 
the calorimetric energy after applying the tau neutrino and muon neutrino selections, respectively at the point $\Delta m^2_{14}$ = 22 e$V^2$ and $\theta_{\mu\tau}$ = 0.175 rad.

\begin{figure}[H]
\centering
\begin{subfigure}{.5\textwidth}
  \centering
 \epsfig{file=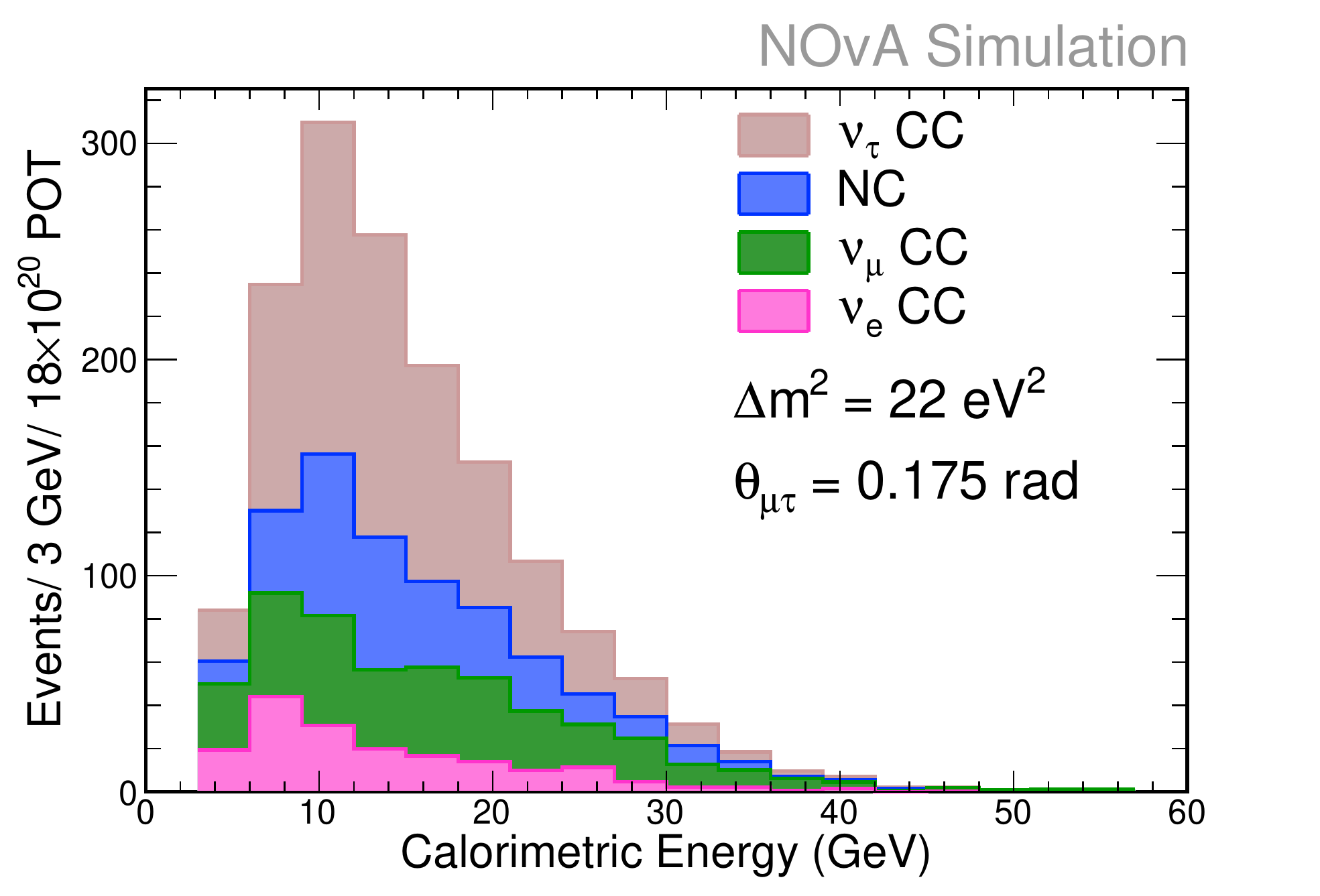,height=2.2in}
   \caption{ $\nu_{\mu} \rightarrow \nu_{\tau}$ selection}
  \label{fig:sub1}
\end{subfigure}%
\begin{subfigure}{.5\textwidth}
  \centering
 \epsfig{file=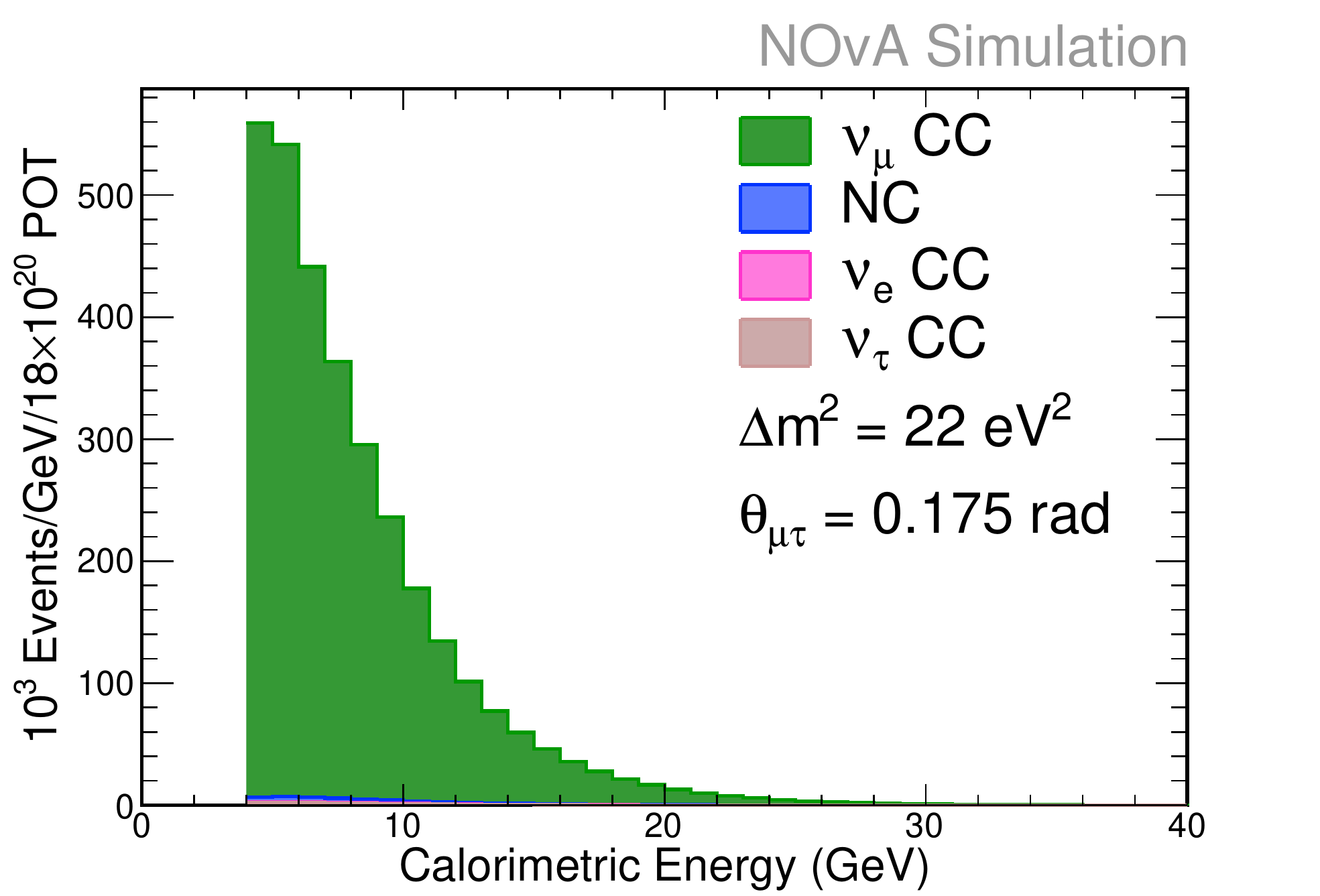,height=2.2in}
  \caption{ $\nu_{\mu} \rightarrow \nu_{\mu}$ selection}
  \label{fig:sub2}
\end{subfigure}
\caption{Simulated calorimetric energy distribution for the selected neutrino events at the point $\Delta m^2_{14}$ = 22 e$V^2$ and $\theta_{\mu\tau}$ = 0.175 rad. scaled to 18$\times10^{20}$ POT.}
\label{fig:test}
\end{figure}
%
%


%
%


\section{Sensitivity to Sterile Neutrinos}
As shown in Fig. 4, NOvA will be competitive with previous experiments [2-9] after 3 years of running(18$\times10^{20}$ POT). Future work will include improving the efficiency for our selection of $\nu_{\tau}$ according to their $\tau$ decay mode.
%
%
\begin{figure}[H]
\begin{center}
\epsfig{file=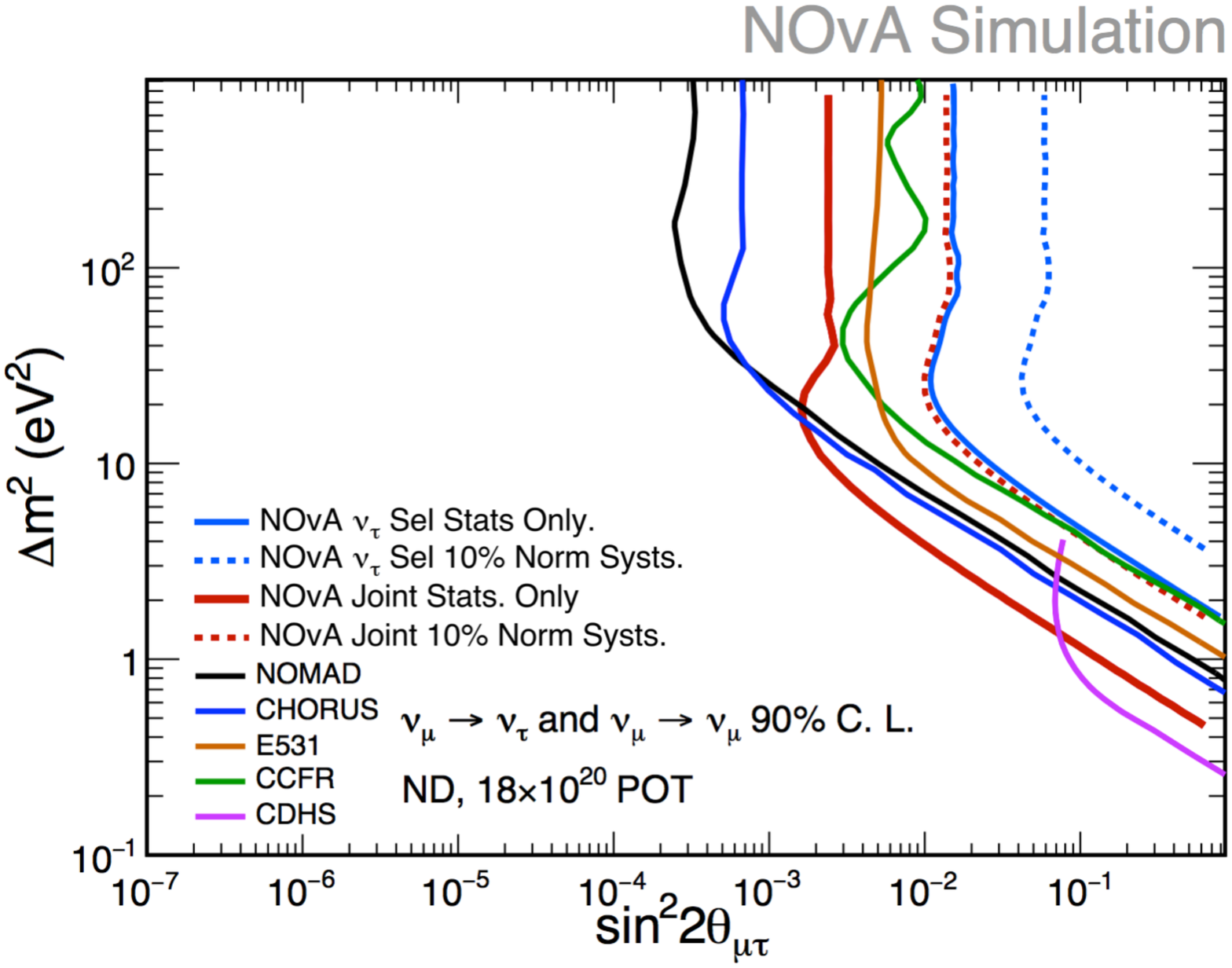,height=3. in}
\caption{90\% C. L. sensitivity to the sterile neutrinos for 18$\times10^{20}$ POT  as compared to previous results from other experiments.}
\label{fig:magnet}
\end{center}
\end{figure}

\section{Acknowledgements}
The NOvA collaboration uses the resources of the Fermi National Accelerator Laboratory (Fermilab), a U.S. Department of Energy, Office of Science, HEP User Facility. Fermilab is managed by Fermi Research Alliance, LLC (FRA), acting under Contract No. DE-AC02-07CH11359. This research was supported by the U.S. Department of Energy; the U.S. National Science Foundation; the Department of Science and Technology, India; the European Research Council; the MSMT CR, GA UK, Czech Republic; the RAS, RMES, and RFBR, Russia; CNPq and FAPEG, Brazil; and the State and University of Minnesota. We are grateful for the contributions of the staff at Fermilab and the NOvA Far Detector Laboratory.

\end{document}